\documentclass[twocolumn,notitlepage,prl]{revtex4-1}

\usepackage{graphicx}
\usepackage{float} 
\usepackage{subfigure} 
\usepackage{dcolumn}
\usepackage{bm}
\usepackage{amsthm}
\usepackage{amsmath}
\usepackage{amssymb}
\usepackage{color}
\usepackage{algorithm}
\usepackage{algorithmicx}
\usepackage{algpseudocode}

\newtheorem{lemma}{Lemma}

\definecolor{Red}{rgb}{1,0,0}
\def\vec#1{{\bm #1}}

\def\ket#1{| #1 \rangle}
\def\bra#1{\langle #1 |}

\def\Tr{\operatorname{Tr}}

\def\poly{\operatorname{poly}}
\def\pe{\operatorname{pe}}
\def\Tr{\operatorname{Tr}}
\def\eigen{\operatorname{eigen}}

\def\O{\mathcal{O}}

\def\R{\mathbb{R}}

\begin{document}
\bibliographystyle{elsarticle-num-names}

\title{Quantum spectral clustering algorithm for unsupervised learning}

\author{Qingyu Li}
\affiliation{Institute of Fundamental and Frontier Sciences, University of Electronic Science and Technology of China, Chengdu, 610051, China}

\author{Yuhan Huang}
\affiliation{Institute of Fundamental and Frontier Sciences, University of Electronic Science and Technology of China, Chengdu, 610051, China}

\author{Shan Jin}
\affiliation{Institute of Fundamental and Frontier Sciences, University of Electronic Science and Technology of China, Chengdu, 610051, China}

\author{Xiaokai Hou}
\affiliation{Institute of Fundamental and Frontier Sciences, University of Electronic Science and Technology of China, Chengdu, 610051, China}

\author{Xiaoting Wang}
\email{xiaoting@uestc.edu.cn}
\affiliation{Institute of Fundamental and Frontier Sciences, University of Electronic Science and Technology of China, Chengdu, 610051, China}

\begin{abstract}
Clustering is one of the most crucial problems in unsupervised learning, and the well-known $k$-means clustering algorithm has been shown to be implementable on a quantum computer with a significant speedup. However, many clustering problems cannot be solved by $k$-means, and a powerful method called spectral clustering is introduced to solve these problems. In this work, we propose a circuit design to implement spectral clustering on a quantum processor with a substantial speedup, by initializing the processor into a maximally entangled state and encoding the data information into an efficiently-simulatable Hamiltonian. Compared with the established quantum $k$-means algorithms, our method does not require a quantum random access memory or a quantum adiabatic process. It relies on an appropriate embedding of quantum phase estimation into Grover's search to gain the quantum speedup. Simulations demonstrate that our method is effective in solving clustering problems and will serve as an important supplement to quantum $k$-means for unsupervised learning. 

\end{abstract}

\maketitle
\begin{figure}
 \centering
 \begin{minipage}[c]{0.2\textwidth}
 \centering
 \includegraphics[width=1\textwidth]{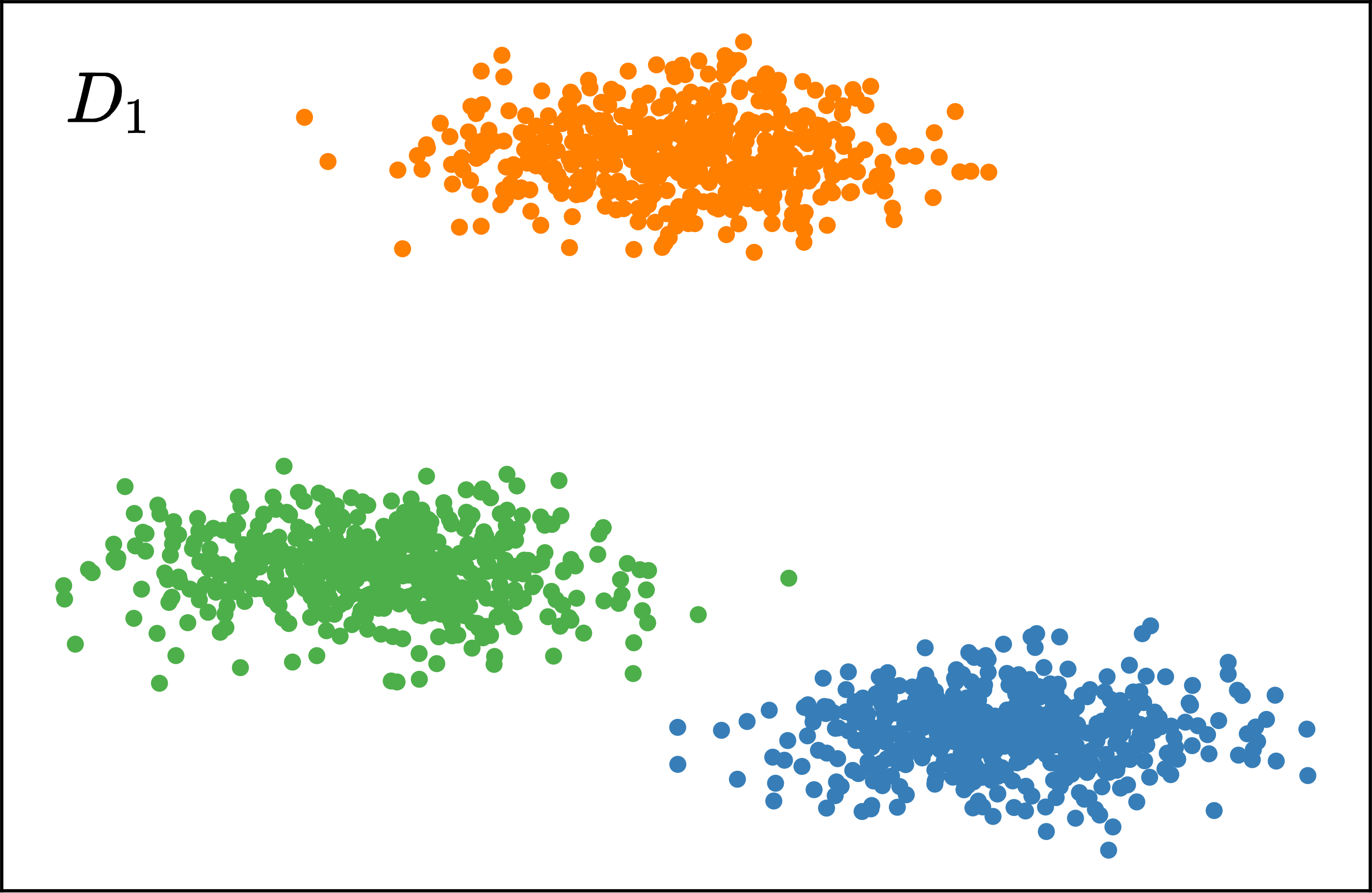}
 {\footnotesize{(a)}}
 \end{minipage}
 \hspace{0.01\textwidth}
 \begin{minipage}[c]{0.2\textwidth}
 \centering
 \includegraphics[width=1\textwidth]{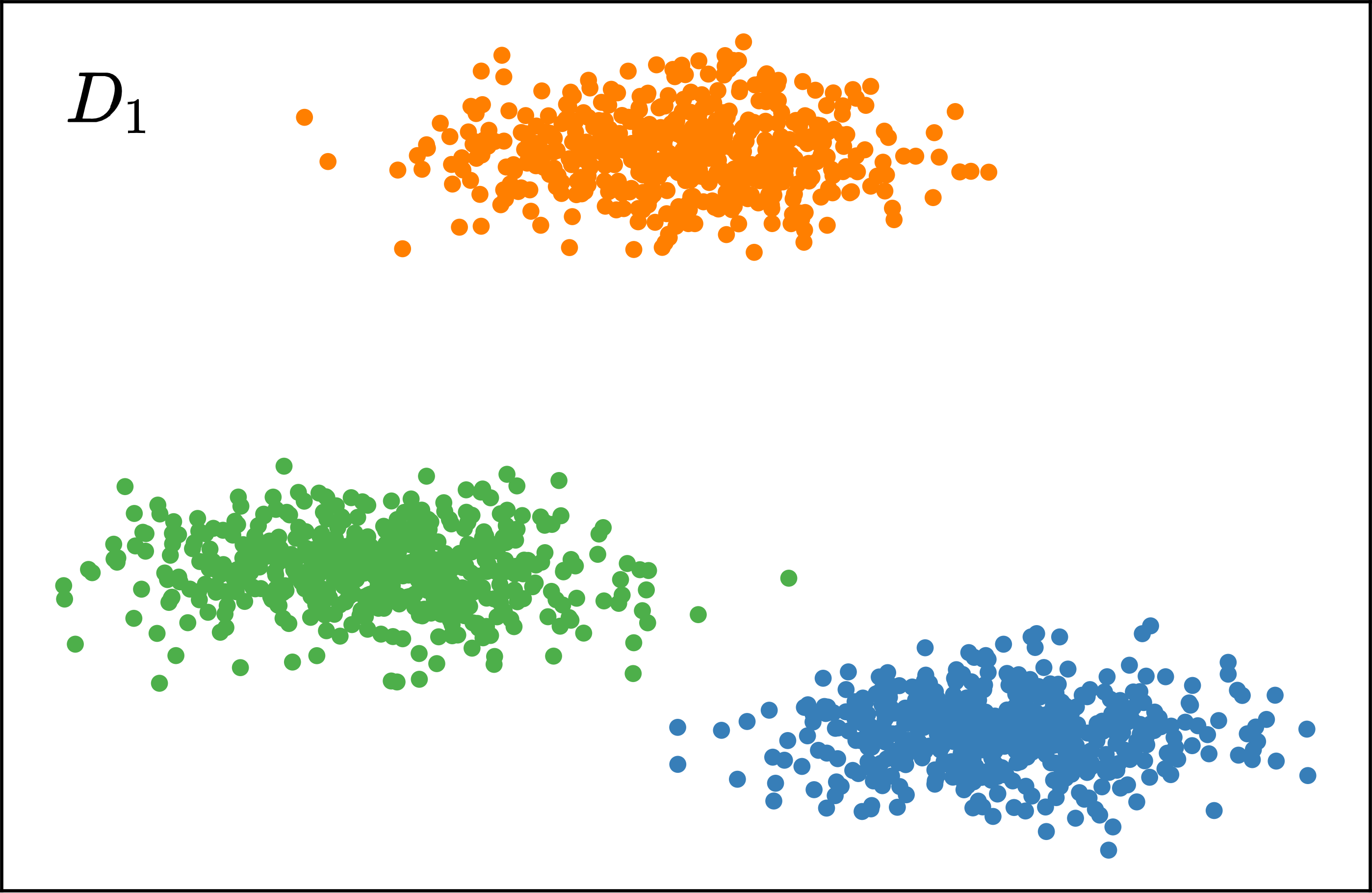}
 {\footnotesize{(b)}}
 \end{minipage}
 \hspace{0.01\textwidth}
 \begin{minipage}[c]{0.2\textwidth}
 \centering
 \includegraphics[width=1\textwidth]{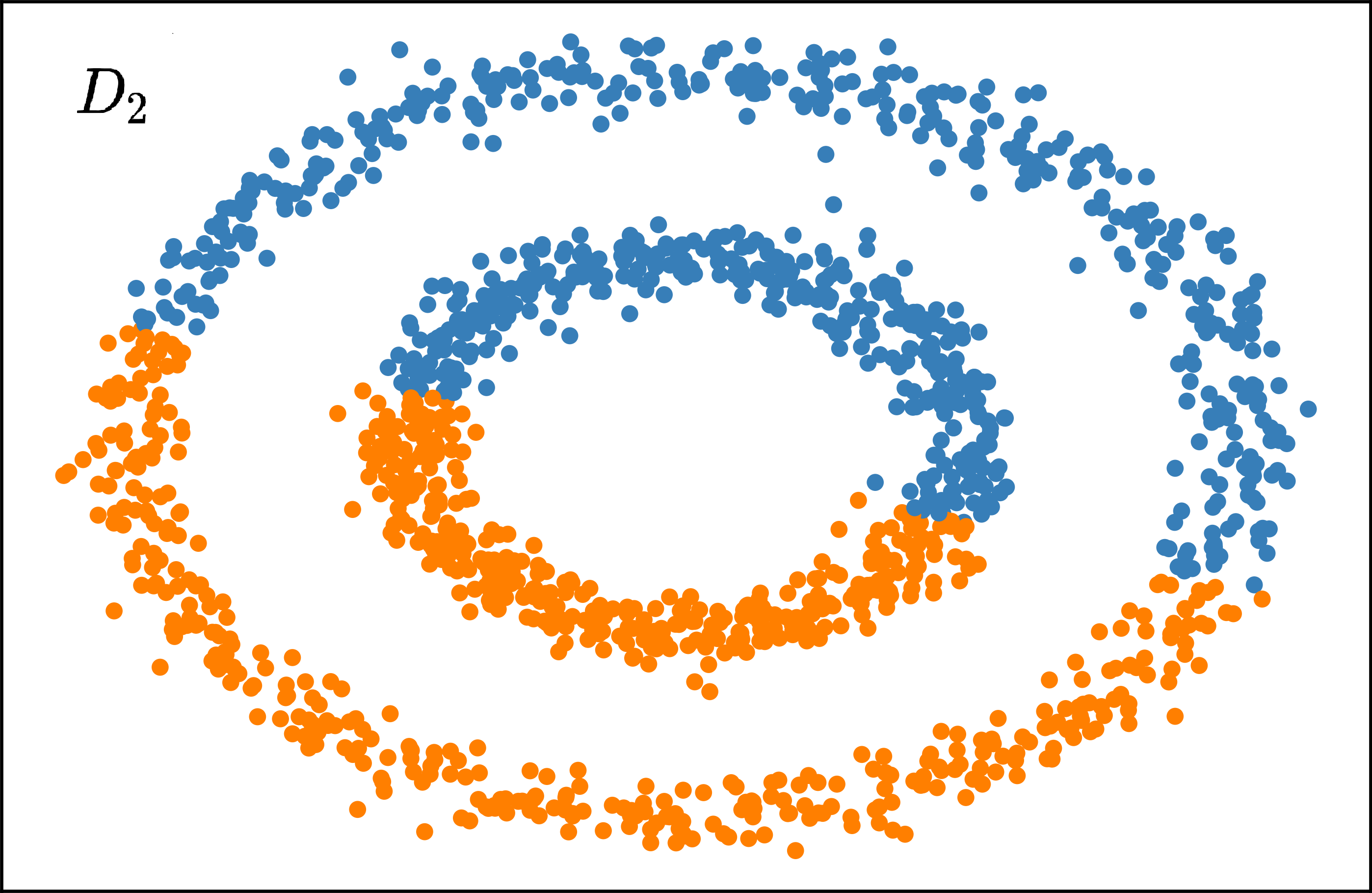}
 {\footnotesize{(c)}}
 \end{minipage}
 \hspace{0.01\textwidth}
 \begin{minipage}[c]{0.2\textwidth}
 \centering
 \includegraphics[width=1\textwidth]{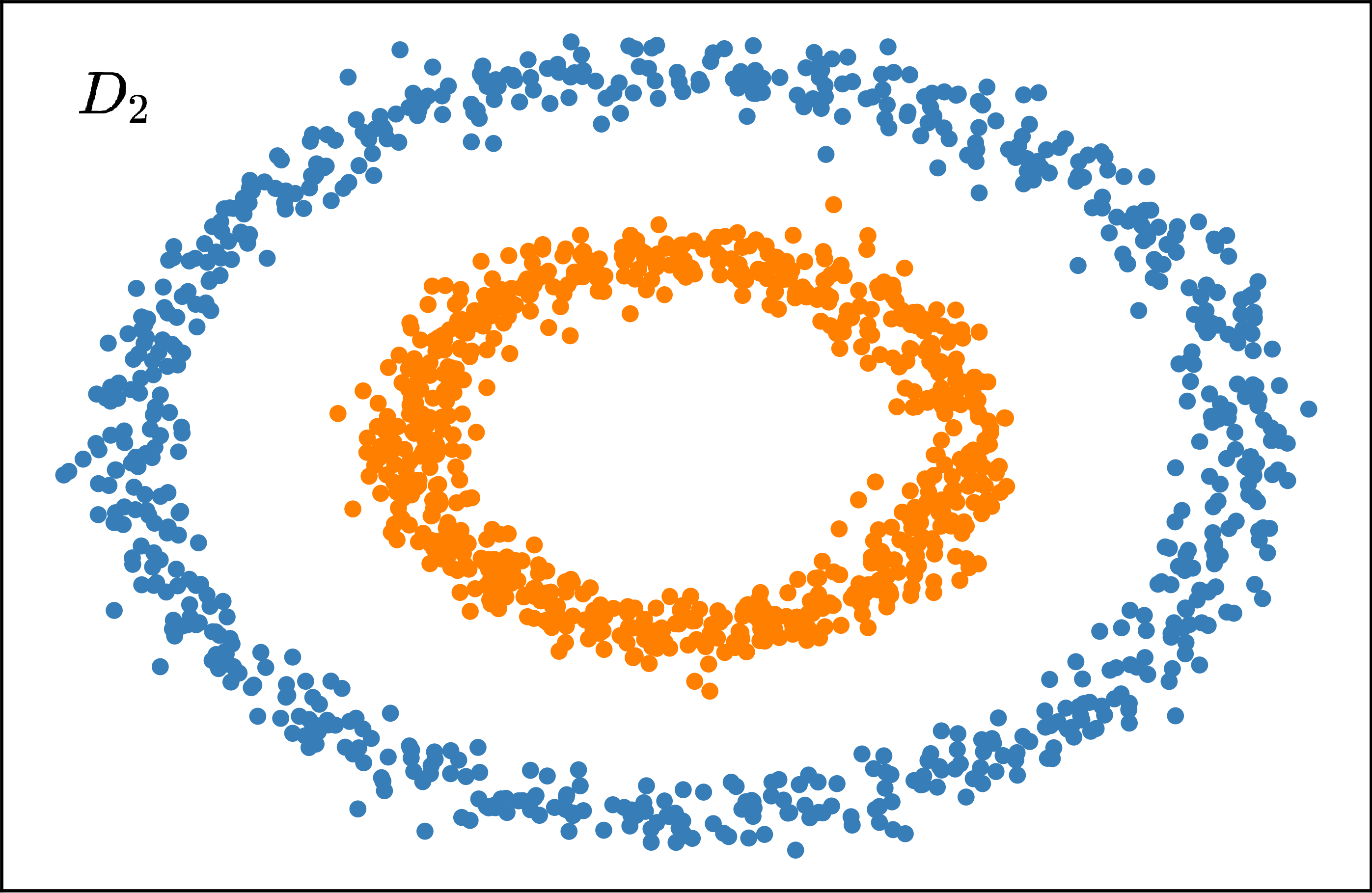}
 {\footnotesize{(d)}}
 \end{minipage}
 \caption{Comparison of clustering results for data sets $D_1$ and $D_2$ under two different methods, $k$-means and spectral clustering:  (a) and (c) under $k$-means, and (b) and (d) under spectral clustering. Both methods give good clustering for $D_1$, but only spectral clustering give a reasonable clustering for $D_2$.}
 \label{Fig.class example}
\end{figure}

Quantum machine learning~(QML) is an interdisciplinary subject connecting quantum computing and machine learning~(ML), with a focus on solving ML problems on quantum processors and obtaining potential quantum speedup or other types of advantage~\cite{Biamonte_QML}. 
Although the exploration of QML is still preliminary, existing results have shown the advantage of many QML algorithms in substantially reducing the computational complexity compared to their classical counterparts. Examples of such advantages include quantum data-fitting~\cite{Wiebe_data_fitting}, quantum support vector machine~(QSVM)~\cite{Rebentrost_QSVM,Li_QSVM}, quantum principal component analysis~\cite{Lloyd_QPCA}, quantum Boltzmann machine~\cite{Amin_QBM}, and quantum reinforcement learning~\cite{Dong_QRL}. Besides supervised learning and reinforcement learning, another important subfield in ML is unsupervised learning, where algorithms are designed to find hidden patterns for a set of unlabeled data. Typical unsupervised learning problems include anomaly detection~\cite{Chandola_Anomaly}, dimensionality reduction~\cite{Maaten_Dimensionality}, and clustering~\cite{Jain_Data_cluster}. Clustering aims to group a set of data points into a number of subgroups based on their similarities, and one of the most popular clustering algorithms is $k$-means. It has been shown that $k$-means algorithm can be implemented on a quantum computer by either converting it to a search problem~\cite{Aimeur_QCA}, or utilizing a quantum random access memory~(QRAM)~\cite{Giovannetti_QRAM} and adiabatic quantum computing~\cite{Farhi_adiabatic} to realize it~\cite{Lloyd_supervisedML}. Depending on the specific design, the quantum $k$-means algorithm can achieve a quadratic or exponential speedup~\cite{Aimeur_QCA,Lloyd_supervisedML}. Nevertheless, not all clustering problems can be solved by $k$-means, and there are important cases of clustering when $k$-means fails, but another method called spectral clustering works, as shown in Figure~\ref{Fig.class example}. Compared with $k$-means, spectral clustering is more flexible and more adaptable to different data distributions~\cite{Ng_SC,Luxburg_SC}. From the view of graph theory, spectral clustering is equivalent to a graph cut problem, which can be solved by calculating the first $k$ smallest eigenvalues and the corresponding eigenvectors of the graph Laplacian matrix. Hence, the essence of a quantum spectral clustering algorithm is to solve an eigenvalue problem on a quantum computer. The first effort to establish a quantum algorithm for spectral clustering is based on biased phase estimation~\cite{Daskin_QSC}, but the success of the algorithm cannot be guaranteed. Another effort is made with a crucial assumption on the availability a QRAM~\cite{Kerenidis_QSC}; in reality, how to physically build a QRAM is still an open question. One major challenge of designing such a quantum algorithm is to create an initial state that overlaps with every eigenvector of the Laplacian matrix with equal probability amplitude. 

In order to address this problem, in this work, we propose an alternative quantum algorithm for spectral clustering based on a bipartite maximally entangled initial state. Compared with existing quantum $k$-means and quantum clustering algorithms, our proposal does not require a QRAM or a quantum adiabatic process. We use a $d$-sparse Hamiltonian to encode the Laplacian matrix of the clustering problem, so that it can be efficiently simulated. Our designed clustering circuit combines Grover's search~\cite{Grover_grover}, quantum phase estimation~\cite{Kitaev_QPE} and $d$-sparse Hamiltonian simulation~\cite{Berry_Hsimulating}, in such a way that the entire circuit complexity has a speedup over the well-known classical eigensolvers. After passing through the clustering circuit, the quantum state will undergo a quantum measurement of a pre-chosen observable. Finally, the measurement outcome is optimized over the choice of the observable, and a desired clustering result will be achieved. Before getting into the detail of our quantum proposal, we will first have a brief review of the classical spectral clustering algorithm. 

\textit{Classical spectral clustering algorithm} --
Given a dataset $D=\{\vec{v}_i\}_{i=0}^{N-1}\subset \mathbb{R}^M$, a clustering task is to group the points in $D$ into $k$ subgroups based on their similarities, and the clustering outcome is described by a partition $\{P_j\}_{j=0}^{k-1}$ with $D=\bigcup_{j=0}^{k-1} P_j$ and $P_i \cap P_j=\varnothing$, for $i\neq j$. The value of $k$ is given as an input of the clustering problem. 

In order to apply the spectral clustering algorithm, we further assume the similarity between points $\vec{v}_i$ and $\vec{v}_j$ is characterized by a similarity function $S$: $S(\vec{v}_i,\vec{v}_j)=S(\vec{v}_j,\vec{v}_i)\in [0,+\infty)$, satisfying $S(\vec{v}_i,\vec{v}_j)$ is large if the two points are from the same subgroup, and is small if they are from different subgroups. Compared with the $k$-means algorithm that deals with the dataset $D$ directly, the spectral clustering algorithm deals with the similarity graph $G(V_D,E_{DS})$, whose nodes are points in $D$, and edges between every pair of points that are $(d-1)$-nearest neighbors to each other. In other words, two points $\vec{v}_i$ and $\vec{v}_j$ in $G$ are connected by an edge $e_{ij}$ with a weight $S(\vec{v}_i,\vec{v}_j)$, if they are $(d-1)$-nearest neighbors to each other. Here, $d$ is a value chosen by the user to characterize the local connectivity of every point to its surrounding neighbors. In most applications, $d$ is independent of $N$ and $d\ll N$, which we will take as assumptions of the clustering problem. Then we can define $W$ as the adjacent matrix of $G(V_D,E_{DS})$, with $w_{ij}=S(\vec{v}_i,\vec{v}_j)$ for each edge $e_{ij}$ in $G$. Then the associated Laplacian matrix $L\in \mathbb{R}^{N\times N}$ for $G(V_D,E_{DS})$ is defined by: 
\begin{eqnarray}
  L_{ij}=
  \left\{
  \begin{aligned}
    -w_{ij},&\quad i\neq j,\\
  \sum_{l=0}^{N-1} w_{i l},&\quad i=j.
  \end{aligned}
  \right.
  \label{lapmat}
\end{eqnarray}
$L$ is symmetric and positive-semidefinite, with eigenvectors $\{u_i\}$ and eigenvalues $\{\lambda_i\}$ satisfying $0=\lambda_0\le\lambda_1\le\cdots\leq\lambda_{N-1}$. Notice that $L$ has at least one zero eigenvalue, i.e. $\lambda_0=0$.

The idea of spectral clustering comes from the property of the Laplacian matrix $L$: the number of zero eigenvalues of $L$ is equal to the number of connected components of the similarity graph. Hence, eigenvalues of $L$ equal to or close to zero will provide useful clustering information. The classical spectral clustering algorithm consists of two steps. In Step 1, we calculate the first $k$ smallest eigenvalues $\{\lambda_i\}$, $i=0,\cdots, k-1$ as well as their corresponding eigenvectors $u_i$, and construct the matrix $A \equiv [\vec{u}_{0},...,\vec{u}_{k-1}]\in\mathbb{R}^{N\times k}$. The rows of $A$ are denoted as $\{\vec{y}_i\}$, $i=0,\cdots, N-1$. In Step 2, we apply the $k$-means clustering algorithm to $\{\vec{y}_i\}$ and group them into $k$ subgroups $\{C_j\}_{j=0}^{k-1}$. It turns out that the clustering $\{C_j\}$ on $\{\vec{y}_i\}$ leads to a good clustering $\{P_j\}$ on $D=\{\vec{v}_i\}$: if $\vec{y}_i$ and $\vec{y}_j$ belong to the same subgroup $C_j$, then $\vec{v}_i$ and $\vec{v}_j$ belong to the same subgroup $P_j$~\cite{Ng_SC,Luxburg_SC}. The entire process of spectral clustering is summarized in Algorithm~\ref{Classical_spectral_clustering}. Since the time complexity of numerically calculating the first $k$ smallest eigenvalues of $L$ is $\O(kN^3)$ (e.g. using the inverse power method~\cite{Wilkinson_AEP}), the time complexity of spectral clustering is at least $\O(kN^3)$. 

\begin{algorithm}[H]
  \caption{Classical spectral clustering algorithm}
  \label{Classical_spectral_clustering}
  \hspace*{\algorithmicindent} \textbf{Input}:  A dataset $D=\{\vec{v}_j\}$, a given value $d$, and the number of clusters $k$. \\
  \hspace*{\algorithmicindent} \textbf{Output}: Clusters $P_1,...,P_k$.
  \begin{algorithmic}[1]
  \State Construct the similarity graph $G(V_{D},E_{DS})$ upon $D$ and $S(\vec{v}_i,\vec{v}_j)$;
  \State Compute the graph Laplacian matrix $L$;
  \State Calculate the first $k$ eigenstates $u_0,...,u_{k-1}$ of L; 
  \State Let $A\equiv [\vec{u}_0,...,\vec{u}_{k-1}]\in\mathbb{R}^{N\times k}$ and denote the $i$-th row of $A$  as $\vec{y}_i\in \mathbb{R}^k$;
  \State Cluster $\vec{y}_{i=0,...,N-1} \in \mathbb{R}^k$ into $k$ different clusters $C_0,...,C_{k-1}$ with $k$-means algorithm;
  \State Generate $P_0,...,P_{k-1}$ by $P_j=\{\vec{v}_i|\vec{y}_i\in C_j\}$.
  \end{algorithmic}
\end{algorithm}

\begin{figure*} 
   \centering 
   \includegraphics[width=1\textwidth]{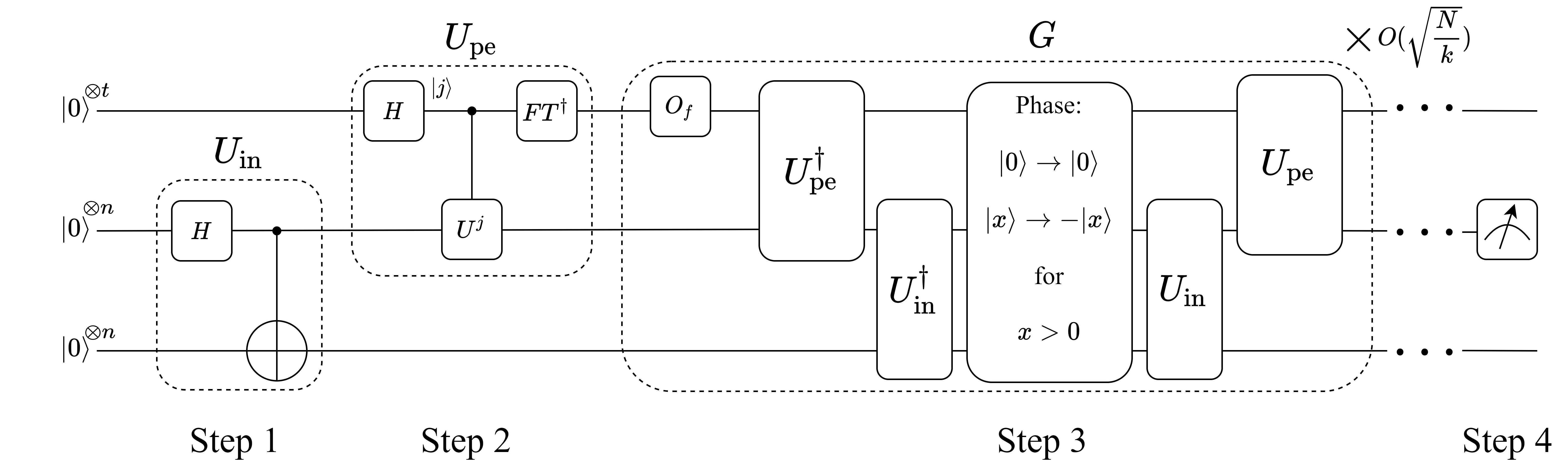} 
   \caption{Quantum circuit illustrating our quantum spectral clustering algorithm. The quantum processor contains three registers, the phase register, the eigenstate register and an ancilla register, with $t$, $n$ and $n$ qubits respectively. The value of $t$ determines the accuracy of phase estimation and $n=\log (N)$. The entire circuit contains four steps: initial state preparation (Step 1), quantum phase estimation with $U=e^{2\pi iL}$ (Step 2), Grover iteration sequence to find the eigenvalues of $L$ less than $\tilde\lambda$ (Step 3), and quantum measurement on the eigenstate register to evaluate and optimize $\langle M\rangle=\Tr(\rho M)$, where $M=XX^T$ and $X$ is the clustering indicator matrix (Step 4). }   
   \label{Fig.quantumFlow} 
\end{figure*}

\textit{Quantum spectral clustering algorithm} --
\label{QSC}
As discussed above, the critical step of spectral clustering is to calculate the eigenvalues of the Laplacian matrix $L$ for data set $D$. 
In this work, we assume that as the set size $N=|D|$ grows, all points in $D$ are confined in the same compact region $D_{\text{c}}\subset \R^M$. 
In the classical case, there are many ways of defining the similarity function $S$, the similarity graph $G(V_{D},E_{DS})$ and the corresponding $L$; in the quantum case, analogous to the classical case, for a given $d$, we can define the $(d-1)$-nearest neighbor graph $G(V_{D},E_{DS})$, with its adjacent matrix $W=(w_{ij})$. For convenience, we choose all nonzero $w_{ij}$ to be $1$. Then the Laplacian matrix $L$ can be defined by~(\ref{lapmat}). Assuming $d\ll N$, we have $L$ is Hermitian and $d$-sparse. We also assume $N=2^n$, so that $L$ can be encoded as a Hamiltonian on an $n$-qubit system. 
In addition, $L$ can be further rescaled by a factor of $\frac{1}{2d}$ so that after rescaling all eigenvalues of $L$ fall into the interval $[0,1]$. 
Such treatment will not alter the corresponding eigenstates, nor the property of the Laplacian matrix, but will make it convenient to encode $L$ into the quantum circuit. 
We denote the eigenvalues and eigenstates of $L$ as $\{\lambda_j\}_{j=0}^{N-1}$ and $\{\ket{u_j}\}_{j=0}^{N-1}$, respectively. In addition, for every $\lambda_j$, its binary representation is $\lambda_j=0.\lambda_{j1}\lambda_{j2}\lambda_{j3}...$, where $\lambda_{jl}\in \{0,1\}$.
It turns out that quantum phase estimation (QPE)~\cite{Kitaev_QPE} embedded in Grover's search is a powerful method to calculate the eigenvalues of $L$. 
For $U=e^{2\pi i L}$ and one eigenpair $(e^{2\pi i \lambda_k},\ket{u_k})$ of $U$, the quantum phase estimation circuit $U_{\text{pe}}$ can be constructed from the inverse quantum Fourier transform and a series of controlled-$U^{j}$ gates, satisfying
\begin{equation}
U_{\text{pe}}|0\rangle^{\otimes t}|u_k\rangle=|\tilde\lambda_k\rangle|u_k\rangle,
\end{equation}
where $|\tilde\lambda_k\rangle=|\lambda_{k1}\lambda_{k2}...\lambda_{kt}\rangle$ and its measurement outcome is a t-bits estimate of $\lambda_k$, up to an error precision $\O(2^{-t})$. For convenience, the first register is called the phase register, and the second called the eigenstate register. To implement quantum spectral clustering, it is sufficient to choose $U=e^{2\pi iL}$ in $U_{\text{pe}}$ and then to calculate the eigenstates of $L$. Notice that the value of $d$ cannot be chosen too large; otherwise it would mistakenly connect two clusters that are generically distinctive. Hence, for our case of clustering on a compact set, $d$ has an upper bound independent of $N$. Thus, $U=e^{2\pi iL}$ can be efficiently simulated on a quantum circuit~\cite{Berry_Hsimulating}. The entire procedure of our quantum algorithm for spectral clustering consists of four steps, as illustrated in Figure~\ref{Fig.quantumFlow}.

\emph{Step 1: initial state preparation}. 

If the input state of $U_{\text{pe}}$ is $\ket{0}|\phi\rangle$, where $\ket\phi=\sum_{i=0}^{N-1}\alpha_i \ket{u_i}$, then we have:
\begin{equation}
  U_{\text{pe}}|0\rangle|\phi\rangle=\sum_{i=0}^{N-1}\alpha_iU_{\text{pe}}|0\rangle|u_i\rangle=\sum_{i=0}^{N-1}\alpha_i|\lambda_i\rangle|u_i\rangle.
  \label{randomstate}
\end{equation}
Since $\{\ket{u_i}\}$ are unknown before the calculation, some of these amplitudes $\alpha_i$ could be significantly small for a poor choice of $\ket\phi$, which could lead to a problem in the subsequent steps. To address this problem, we propose to introduce an ancilla register and prepare a bipartite maximally entangled initial state. We need the following lemma for further analysis\cite{Poulin_gibbs}. 

\begin{lemma}
  \label{lemma1}
  For an $N$-dimensional quantum system, let $\{|i\rangle\}$ be the computational basis, and $\{|v_j\rangle\}$ be another orthonormal basis arbitrarily chosen, $j=0,1,\cdots,N-1$. Then we have
  \begin{equation}
    \ket{\Phi}\equiv\frac{1}{\sqrt{N}}\sum_{i=0}^{N-1}|i\rangle|i\rangle=\frac{1}{\sqrt{N}}\sum_{j=0}^{N-1}|v_j\rangle|v_j^\ast\rangle
      \label{maximally entangled state}
  \end{equation}
  where $|v_j^\ast\rangle$ is the complex conjugate of $|v_j\rangle$.
  \end{lemma}

  \begin{proof}
    Assuming $|v_j\rangle=\sum_i\alpha_{ji}|i\rangle$, since
    \begin{align*}
        |i\rangle = \sum_j|v_j\rangle\langle v_j|i\rangle=\sum_{j,k,m}\alpha_{jk}\alpha_{jm}^\ast|k\rangle\langle m|i\rangle= \sum_{j,k}\alpha_{jk}\alpha_{ji}^\ast|k\rangle,
    \end{align*}
    we have:
    \begin{align*}
     \begin{split}
        \sum_j|v_j\rangle|v_j^\ast\rangle&=\sum_j\sum_k\alpha_{jk}|k\rangle\sum_i\alpha_{ji}^\ast|i\rangle\\
        &=\sum_i\big(\sum_{j,k}\alpha_{jk}\alpha_{ji}^\ast|k\rangle\big)|i\rangle=\sum_{i}|i\rangle|i\rangle.
      \end{split}
    \end{align*}
    \end{proof}

This lemma implies that the maximally entangled state $\ket{\Phi}$ can be expressed in any orthonormal basis, either known or unknown. In particular, we can choose $\ket{v_i}=\ket{u_i}$ in (\ref{maximally entangled state}), where $\{\ket {u_i}\}_{i=0}^{N-1}$ is the eigenbasis of $L$ and is unknown before the calculation. The significance of this lemma is, $\ket{\Phi}$ can be efficiently prepared using the known computational basis $\{|i\rangle\}$, and then it can be expressed in terms of the unknown basis $\{\ket{u_i}\}$, but with known coefficients. Notice that $\ket{\Phi}$ can be efficiently constructed using $U_{\text{in}}\equiv \prod_{i=1}^{n}\text{CNOT}_{i,i+n}H^{\otimes n}$: $\ket{\Phi}= U_{\text{in}}\ket{0}^{\otimes 2n}$. Here, $U_{\text{in}}$ consists of $n$ Hadamard gates and $n$ CNOT gates, where $H^{\otimes n}$ acts on the first $n$ qubits and $\text{CNOT}_{(i,i+n)}$ denotes the $\text{CNOT}$ gate with the $i$-th qubit as the control and the $(i+n)$-th qubit as the target, as illustrated in Figure~\ref{Fig.quantumFlow}. Hence, in Step 1, we prepare the three registers, the phase register, the eigenstate register, and the ancilla register into the initial state $\ket{\psi_0}\equiv \ket{0}^{\otimes t}\ket{0}^{\otimes n}\ket{0}^{\otimes n}$, and then apply $U_{\text{in}}$ to get: $U_\text{in}\ket{\psi_0}=\frac{1}{\sqrt{N}}\sum_{i=0}^{N-1}|0\rangle^{\otimes t}\ket{i}\ket{i}$.

\emph{Step 2: applying quantum phase estimation}. 

After applying $U_\text{in}$ to $\ket{\psi_0}$, we apply the phase estimation circuit $U_{\text{pe}}$ to obtain:
\begin{equation}
  \begin{split}
    \ket{\psi}_{\pe}\equiv U_{\text{pe}}U_\text{in}\ket{\psi_0}&=\frac{1}{\sqrt{N}}\sum_{i=0}^{N-1}U_{\text{pe}}|0\rangle^{\otimes t}\ket{i}\ket{i}\\
    &=\frac{1}{\sqrt{N}}\sum_{i=0}^{N-1}\ket{\lambda_i}\ket{u_i}\ket{u_i^*},
  \end{split}
    \label{pestate}
\end{equation}
where we have used $U_{\text{pe}}|0\rangle^{\otimes t}|u_i\rangle=\ket{\lambda_i}\ket{u_i}$. 
From this, we can see the advantage of preparing $\ket{\Phi}$: each overlap $\alpha_i$ between $|\psi\rangle_{\text{pe}}$ and every $\ket{\lambda_i}\ket{u_i}\ket{u_i^*}$ is known and equal to $\frac{1}{\sqrt{N}}$, even though $\{\ket{u_i}\}$ is unknown. 

In order to efficiently construct $U_{\pe}$, we need to efficiently construct the following controlled-$U$ gate, with $U=e^{2\pi iL}$: 
\begin{align}
CU= \ket{0}\bra{0}\otimes I +\ket{1}\bra{1}\otimes e^{2\pi iL}= e^{2\pi i(\ket{1}\bra{1}\otimes L)},
\end{align}
where $H\equiv \ket{1}\bra{1}\otimes L $ is also a $d$-sparse matrix. According to the well-known results in quantum simulation, since $H=\ket{1}\bra{1}\otimes L $ is $d$-sparse, $CU$ and hence $U_{\pe}$ can be efficiently constructed, and the quantum complexity for $U_{\pe}$ is $\O(\poly(\log N)d^4/\epsilon)$, where $\epsilon$ represents the error of the estimated phase~\cite{Berry_Hsimulating}

\emph{Step 3: applying Grover's search to find the eigenvalues of $L$ less than the threshold.} 

In classical spectral clustering, the number of clusters $k_0$ is chosen as the input of the clustering problem, and a classical eigenvalue-solving algorithm will be applied to find the $k_0$ smallest eigenvalues of $L$. In comparison, in our quantum spectral clustering algorithm, we choose a threshold value $\tilde{\lambda}>0$, and apply Grover's search~\cite{Grover_grover} to $|\psi\rangle_{\text{pe}}$ to find eigenstates of $L$ smaller than $\tilde{\lambda}$. Specifically, given $\tilde{\lambda}$, the desired classical oracle $f$ can be defined as follows:
\begin{eqnarray}
  f(x)=
  \left\{
  \begin{aligned}
    1,\ \frac{x}{2^t} < \tilde{\lambda} \\   
    0,\  \frac{x}{2^t} \ge \tilde{\lambda}
  \end{aligned}
  \right.
  \label{Of}
\end{eqnarray}
where $x$ is a $t$-bit Boolean variable. Then based on the reversible classical circuit that generates $f$, we can construct the corresponding quantum circuit that generates the quantum oracle $O_f$~\cite{Watrous}, satisfying 
\begin{eqnarray}
O_f \ket{x}=(-1)^{f(x)} \ket{x}.
\end{eqnarray}
Such $O_f$ will add a phase $-1$ to all $\ket{x}$ satisfying $\frac{x}{2^t}<\tilde{\lambda}$. Then based on $O_f$, we can construct the Grover iteration $G$,
\begin{equation}
    G\equiv U_{\text{inv}}O_f=(2|\psi\rangle_{\text{pe}}\langle\psi|_{\text{pe}}-I)O_f,
\end{equation}
where $U_{\text{inv}}\equiv (2|\psi\rangle_{\text{pe}}\langle\psi|_{\text{pe}}-I)$ is the initial-state inversion with respect to $\ket{\psi}_{\pe}$. Hence, the constructions of $f$ and $O_f$ are completely determined by the value of $\tilde{\lambda}$, and does not require any prior information about the eigenvalues of $L$. 

In addition, a well-known result in quantum computation is, if the circuit complexity to generate the classical oracle $f$ is $C$, then the circuit complexity of the quantum oracle $O_f$ is $\O(\poly(C))$~\cite{Watrous}. For our case, since division and comparison are performed on $t$ bits to evaluate $f$, the classical circuit complexity to calculate $f$ in (\ref{Of}) is $\poly(t)$, and hence the complexity of the above $O_f$ is $\O(\poly(t))$, where $t=n+\lceil 2+\log \frac{1}{2\epsilon_0}\rceil$, $n=\lceil \log N \rceil$ and $1-\epsilon_0$ is the success probability of phase estimation. Hence, the circuit complexity to construct $O_f$ is $\O\big(\poly(\log(N))\big)$, i.e., $O_f$ and hence $G$ can be efficiently constructed.

Since we are only interested in finding eigenvalues of $L$ close to zero, $\tilde{\lambda}$ should be sufficiently small. As a rule of thumb, $\tilde{\lambda}=\frac{1}{2N}$ is a good first trial, and will give a reasonably good clustering. The value of $\tilde{\lambda}$ determines the number $k$ of eigenvalues smaller than $\tilde{\lambda}$, and we can rewrite $\ket{\psi}_{\pe}$ as: 
\begin{equation}
 \label{psi_pe}
    |\psi\rangle_{\text{pe}}=\frac{1}{\sqrt{N}}\sum_{\lambda_i<\tilde{\lambda}}|\lambda_i\rangle|u_i\rangle|u_i^\ast\rangle+ \frac{1}{\sqrt{N}}\sum_{\lambda_i\geq\tilde{\lambda}}|\lambda_i\rangle|u_i\rangle|u_i^\ast\rangle.
\end{equation}
After applying a standard Grover iteration sequence of length $r=\Big\lceil\frac{\pi}{4}\sqrt{\frac{N}{k}}\Big\rceil$ to the initial state $\ket{\psi}_{\pe}$, we obtain the output state:
\begin{equation}
    \label{outputstate}
    |\psi\rangle_{\text{out}}=\frac{1}{\sqrt{k}}\sum_{\lambda_i< \tilde{\lambda}}|\lambda_i\rangle|u_i\rangle|u_i^\ast\rangle,
\end{equation}
with high probability as long as $k\ll N$. Alternatively, we can apply a fixed-point Grover's search~\cite{Luo_Fixed_p} to $\ket{\psi}_{\pe}$ with the final state converging to $|\psi\rangle_{\text{out}}$ with an arbitrarily high precision. The total quantum complexity in Step 3 in terms of $N$ and $k$ is $\O(\sqrt{\frac{N}{k}}\poly(\log N))$; in comparison, on a classical computer, the time complexity is $\O(kN^3)$ using the inverse power method~\cite{Wilkinson_AEP}.

At this point, before we move on to Step 4, we need to find the actual value of $k$, given $\tilde \lambda$. This can be achieved by applying the quantum counting circuit to $\ket{\psi}_{\pe}$~\cite{Brassard_QCounting}. The value of $k$ will then be used to construct the clustering indicator matrix $X$ in Step 4. Details of the quantum counting circuit will be discussed in the next section.

\emph{Step 4: taking quantum measurement and optimizing over the clustering indicator matrix} 

At the end of Step 3, after we obtain the value of $k$ using quantum counting, we need to repeat the quantum circuit of $U_{\text{in}}$, $U_{\pe}$ and the Grover's iterations to prepare the register into $\ket{\psi}_{\text{out}}$ again. Then we take a quantum measurement on the eigenstate register, whose density matrix can be derived by taking the partial trace of $\ket{\psi}_{\text{out}}$:
\begin{eqnarray}
 \rho=\Tr_{1,3}\big( \ket{\psi}_{\text{out}}\bra{\psi}_{\text{out}} \big)=\frac{1}{k}\sum_{i=0}^{k-1}|u_i\rangle\langle u_i|=\frac{1}{k}AA^T
\end{eqnarray}
where $A\equiv [\vec{u}_0,\ldots,\vec{u}_{k-1}]\in\mathbb{R}^{N\times k}$, with rows of $A$ denoted as $\{\vec y_j\}$. As discussed in the classical spectral clustering algorithm, if we apply a clustering for $\{\vec y_j\}$, then the clustering outcome for $\{\vec y_j\}$ will correspond to a good clustering outcome for $\{\vec v_j\}$. In addition, the clustering for $\{\vec y_j\}$ can be conveniently realized in our proposal due to the relation $\rho=\frac{1}{k}AA^T$. Specifically, for any $k$-partition $\{C_i\}_{i=0}^{k-1}$ of $\{\vec{y}_i\}_{i=0}^{N-1}$, we define the clustering indicator matrix $X=(x_{ij})=[\vec{x}_0,\ldots, \vec x_{k-1}]\in\mathbb{R}^{N\times k}$ satisfying
\begin{eqnarray}
  x_{ij}=
  \left\{
  \begin{aligned}
    \frac{1}{\sqrt{s_j}},\ \vec{y}_i\in C_j \\  
    0,\ \vec{y}_i\notin C_j
  \end{aligned}
  \right.
  \label{Kmeans}
\end{eqnarray}
where $s_j=|C_j|$. Then the clustering of $\{\vec{y}_i\}_{i=0}^{N-1}$ can be formulated as an optimization problem with the following objective function~\cite{Zha_Srk}: 
\begin{equation}
  \begin{split}
    \max_{X}&\frac{1}{k}\sum_{i=0}^{k-1}\Tr(\vec{u_i}\vec{u_i}^TXX^T)=\max_{X} \Tr(\rho XX^T)\\
    &=\max_{X} \Tr(\rho M)=\max_{X} \ \langle M\rangle,
  \end{split}
    \label{maxtrace}
\end{equation}
subject to $X$ defined by~(\ref{Kmeans}), where $M\equiv XX^T$ can be considered as a measurement observable. In our proposal, the above optimization problem is equivalent to finding an optimal observable $M=XX^T$ satisfying~(\ref{Kmeans}) to maximize $\langle M\rangle =\Tr(\rho M)$. Strictly speaking, finding the exact optimal $X$ is an NP-hard problem, but for most applications of clustering, a sub-optimal solution is sufficient. Many heuristic algorithms, including hill climbing~\cite{Russell_AI}, are good enough to find an acceptable sub-optimal $X$ within a polynomial number $\O(kN)$ of iterations. We denote $C_{\eigen}$ as the quantum eigenvalue-solving circuit composed by $U_{\text{in}}$, $U_{\pe}$ and the Grover's iteration sequence, and it maps $\ket{\psi_0}$ to $\ket{\psi}_{\text{out}}$. In each iteration of the hill climbing algorithm, the quantum circuit $C_{\eigen}$ is repeatedly implemented for $n_M$ times to get an estimation of $\langle M\rangle$. Since the complexity of $C_{\eigen}$ is $\O(\sqrt{N/k}\poly(\log N))$, the total quantum circuit complexity for $C_{\eigen}$ in all these $\O(kN)$ iterations becomes $\O(\sqrt{k}N^{3/2}\poly(\log N))$, where we have used the fact that $n_M$ only depends on the estimation accuracy of $\langle M\rangle$, and does not increase as $N$ increases. 

\textit{Determine the number of eigenvalues below threshold} --
\begin{figure} 
  \centering 
  \includegraphics[width=0.5\textwidth]{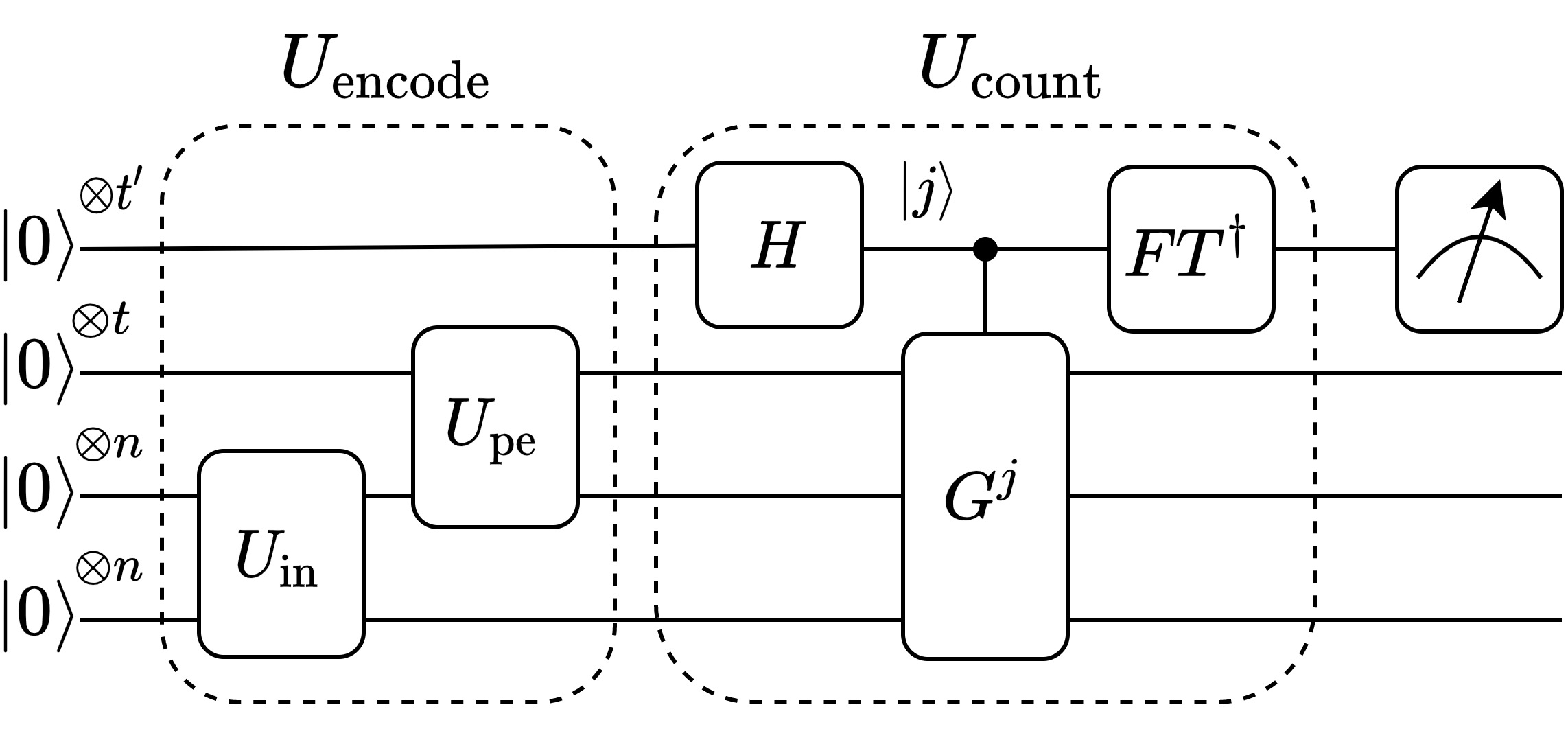} 
  \caption{
    Schematic circuit for quantum counting to estimate calculate the number $k$ of eigenvalues of $L$ less than $\tilde\lambda$, where $U_\text{encode}\equiv U_{\text{pe}}U_\text{in}$, and $G\equiv U_{\text{inv}}O_f$. 
    }   
  \label{Fig.quantumcounting} 
\end{figure}

As mentioned above, at the end of Step 3 in Figure~\ref{Fig.quantumFlow}, based on $\ket{\psi}_{\text{pe}}$, we need to calculate the value of $k$, i.e., the number of eigenvalues (with multiplicities) of $L$ smaller than the given threshold $\tilde{\lambda}$. Specifically, from~(\ref{psi_pe}), we have:
\begin{align*}
    \ket{\psi}_{\text{pe}}&\equiv \sqrt{\frac{N-k}{N}}\ket{\alpha}+\sqrt{\frac{k}{N}}\ket{\beta}=\cos\frac{\theta}{2}\ket{\alpha}+\sin\frac{\theta}{2}\ket{\beta},\\
    \ket{\alpha}&\equiv \sqrt{\frac{1}{N-k}}\sum_{\lambda_i\geq\tilde{\lambda}}\ket{\lambda_i}\ket{u_i}\ket{u_i^\ast},\\
    \ket{\beta}&\equiv\sqrt{\frac{1}{k}}\sum_{\lambda_i<\tilde{\lambda}}\ket{\lambda_i}\ket{u_i}\ket{u_i^\ast}.
\end{align*}
Hence, $\ket{\psi}_{\text{pe}}$ is located on a 2-dimensional subspace, on which $G$ is invariant and has the form: 
\begin{eqnarray}
G= \begin{pmatrix}
\cos\theta & -\sin\theta \\
\sin\theta & \cos\theta
\end{pmatrix}.
\end{eqnarray}
$G$ has two eigenvalues $\mu_1=e^{i\theta}$ and $\mu_2=e^{i(2\pi-\theta)}$, with two corresponding eigenvectors $\ket{a_1}$ and $\ket{a_2}$. 
For $k\ll N$, we have $0< \theta=2\arcsin\sqrt{\frac{k}{N}} < \frac{\pi}{2}$. We can apply the quantum counting circuit $U_\text{count}$ in Figure~\ref{Fig.quantumcounting} to find the value of $\theta$. Specifically, for input state $\ket{0}^{\otimes t'}\ket{\psi}_\text{pe}$, the output state of $U_\text{count}$ becomes:
\begin{equation}
  U_\text{count}\ket{0}^{\otimes t'}\ket{\psi}_\text{pe}=\alpha_1\ket{\theta}\ket{a_1}+\alpha_2\ket{2\pi-\theta}\ket{a_2},
\end{equation}
Finally, through measuring the phase register, we obtain a measurement outcome of either $\theta$ or $2\pi-\theta$. We can tell which is which since the former is smaller than $\frac{\pi}{2}$ and the latter is larger than $\frac{3\pi}{2}$. Thus, we find the value of $\theta$, and so as the value of $k$. By increasing the number of qubits in the phase register, one can reduce the uncertainty $|\Delta k|$ to a value less than $1$, and then the value of $k$ can be uniquely determined. The total circuit complexity of quantum counting is $\O(\sqrt{kN})$. 

\begin{algorithm}[H]
  \caption{Quantum spectral clustering algorithm}
  \label{Quantum_spectral_clustering}
  \hspace*{\algorithmicindent} \textbf{Input}: a data set $D=\{\vec{v_i}\}$, a given value $d$, and a threshold $\tilde{\lambda}$. \\
  \hspace*{\algorithmicindent} \textbf{Output}: clusters $P_0,...,P_{k-1}$.
  \begin{algorithmic}[1]
    \State Given $d$, construct the Laplacian matrix $L$;
    \State Apply $U_{\text{in}}$ to $\ket{\psi_0}$ to prepare the state $\frac{1}{\sqrt{N}}\ket{0}\sum_{i=0}^{N-1}\ket{i}\ket{i}=\frac{1}{\sqrt{N}}\ket{0}\sum_{i=0}^{N-1}|u_i\rangle|u_i^\ast\rangle$;
    \State Apply quantum phase estimation $U_{\text{pe}}$ to generate $\ket{\psi}_\text{pe}=\frac{1}{\sqrt{N}}\sum_{i=0}^{N-1}|\lambda_i\rangle|u_i\rangle|u_i^\ast\rangle$;
    \State Apply the quantum phase estimation $U_{\text{pe}}$ to generate $\ket{\psi}_\text{pe}=\frac{1}{\sqrt{N}}\sum_{i=0}^{N-1}|\lambda_i\rangle|u_i\rangle|u_i^\ast\rangle$;
    \State Apply quantum counting to $\ket{\psi}_\text{pe}$ to find the value of $k$, i.e., the number of eigenvalues of $L$ smaller than $\tilde{\lambda}$;
    \State Repeat Lines 2-3 to get $\ket{\psi}_\text{pe}$, and apply Grover's search to obtain $\rho=\frac{1}{k}\sum_{i=0}^{k-1}|u_i\rangle\langle u_i|$ on the eigenstate register;
    \State Construct an observable $M\equiv XX^T$ with $X$ satisfying~(\ref{Kmeans}), and calculate $\langle M\rangle=\Tr(M\rho)$ through measurement;
    \State Repeat Lines 5-6 and apply hill-climbing algorithm to optimize $\langle M\rangle$ over $X$. The sub-optimal $X=X^*$ gives the desired clustering outcome $P_0,...,P_{k-1}$, with $P_j=\{\vec{v}_i|X^*_{ij}\neq 0\}$.
  \end{algorithmic}
\end{algorithm}

\textit{Complexity analysis} --
In classical spectral clustering, given $L$, the number of clusters $k_0$ is chosen by the user and can be considered as part of the clustering problem input. In comparison, in our quantum spectral clustering algorithm, before implementing the quantum circuit, a threshold $\tilde\lambda$ is chosen, and it completely determines the number $k$ of eigenvalues of $L$ smaller than $\tilde\lambda$, where $k$ is equal to the number of clusters in the final clustering outcome. Hence, given $L$, in our proposal, the value of $\tilde\lambda$ completely determines the final clustering outcome, and hence it is taken as an input of the clustering problem. However, if someone is interested in using our algorithm to group the data set into exactly $k_0$ clusters, then we can do as follows: assuming the eigenvalues of $L$ are sorted in a non-decreasing order, with $0= \lambda_0\le \cdots \le \lambda_{k_0-1}<\lambda_{k_0}\le \cdots \le \lambda_{N-1}\le 1$. Here, we can reasonably assume $\lambda_{k_0}-\lambda_{k_0-1}=\delta>0$, for if $\delta=0$ then the data should be partitioned into $k+1$ rather than $k$ clusters. Then we can apply the binary search algorithm to generate a sequence of $\{\tilde\lambda^{(i)}\}_{i=1}^{m}$ such that the final value $\tilde\lambda^{(m)}\in(\lambda_{k_0-1}, \lambda_{k_0}]$ with exactly $k=k_0$ eigenvalues of $L$ smaller than $\tilde\lambda^{(m)}$. The total number of binary-search iterations is $m=\O(\log\frac{1}{\delta})$. In practice, if $k_0$ corresponds to a good clustering outcome, then $\delta$ must be far from zero, and $m$ must be pretty small. Hence, due to the logarithm property of $m=\O(\log\frac{1}{\delta})$, we can efficiently find the value $\tilde\lambda$ satisfying $k=k_0$ through binary search. 

In addition, similar to $\tilde\lambda$, the value of $d$ is also taken as an input of the clustering problem. The choice of $d$ and the generation of $L$ belong to the data preprocessing part, and are taken as the preliminary information that will be used for the classical and the quantum spectral clustering algorithms. However, $d$ does appear in the circuit complexity expression of simulating the unitary $U=e^{2\pi i L}$. We will take $d$ as a value chosen by the user for the clustering problem, while $N$ and $d$ are taken as independent. In most applications, $d$ is chosen to be a constant much smaller than $N$. In the following, for given values of $d$ and $\tilde\lambda$, we analyze the quantum circuit complexity of all steps in Figure~\ref{Fig.quantumFlow}, including quantum counting. The entire algorithm is summarized in Algorithm~\ref{Quantum_spectral_clustering}. First, quantum counting has a query complexity of $\O(\sqrt{kN})$ and each query has a circuit complexity of $O\big(\poly(\log N)d^4/\epsilon\big)$, where $\epsilon$ denotes the error of the estimated phase in $U_{\pe}$. Then we analyze the four steps of the circuit illustrated in Figure~\ref{Fig.quantumFlow}. The complexity of $U_{\text{in}}$ is $\O(\log N)$, and the complexity of $U_{\pe}$ is $\O(\poly(\log N)d^4/\epsilon)$. For Step 3, the query complexity of the Grover iteration circuit is $\O(\sqrt{\frac{N}{k}})$, and each query has circuit complexity $O\big(\poly(\log N)d^4/\epsilon\big)$. For Step 4, the hill-climbing algorithm consists of $\O(kN)$ iterations, and each iteration repeatedly implements $C_{\eigen}$ for $n_M$ times, with $n_M$ independent of $N$. The total quantum circuit complexity for $C_{\eigen}$ in all these $\O(kN)$ iterations adds up to $\O\big(\sqrt{k}N^{3/2}\poly(\log N)d^4/\epsilon\big)$. Hence, the total complexity of the entire algorithm becomes $\O\big( (kN\sqrt{\frac{N}{k}}+\sqrt{kN})\poly(\log N)d^4/\epsilon \big)=O\big(\sqrt{k}N^{\frac{3}{2}}\poly(\log N)d^4/\epsilon \big)$, demonstrating a notable speedup compared to the complexity of the classical spectral clustering algorithm, e.g., $\O(kN^3)$ for inverse power method to find the $k$ eigenvalues classically~\cite{Wilkinson_AEP}. 

It is worthwhile to point it out that how to choose appropriate values of $d$ and $\tilde \lambda$ for a given clustering problem is an interesting open question in spectral clustering research, but our work focuses on how to implement the spectral clustering through a quantum circuit, and hence both $d$ and $\tilde \lambda$ are taken as given inputs of the clustering problem.

\textit{Numerical simulation} --
\begin{figure*}
  \centering
  \subfigure[]{
  \label{Fig.Result1.1}
  \includegraphics[width=0.32\textwidth]{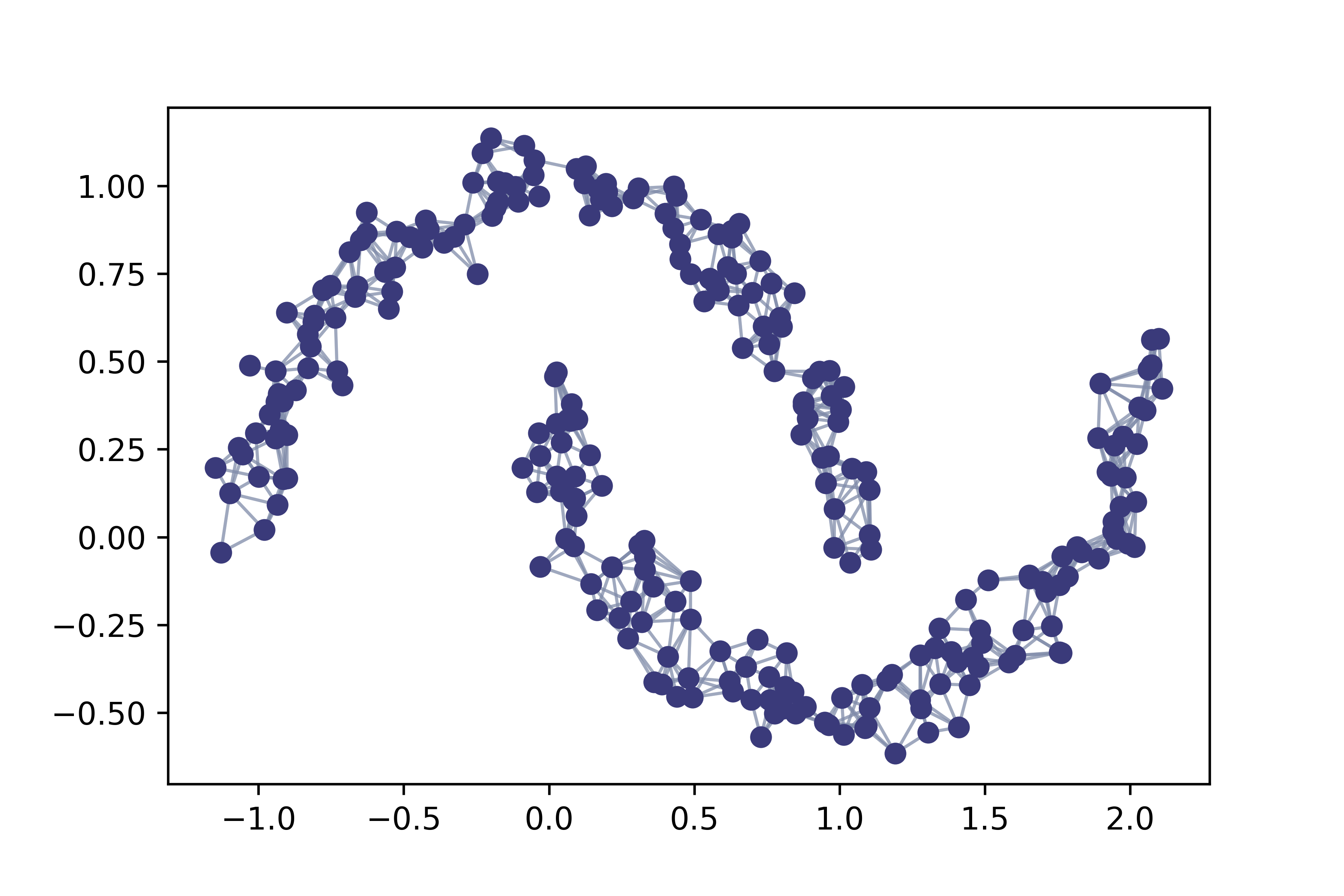}}
  \subfigure[]{
  \label{Fig.Result1.2}
  \includegraphics[width=0.32\textwidth]{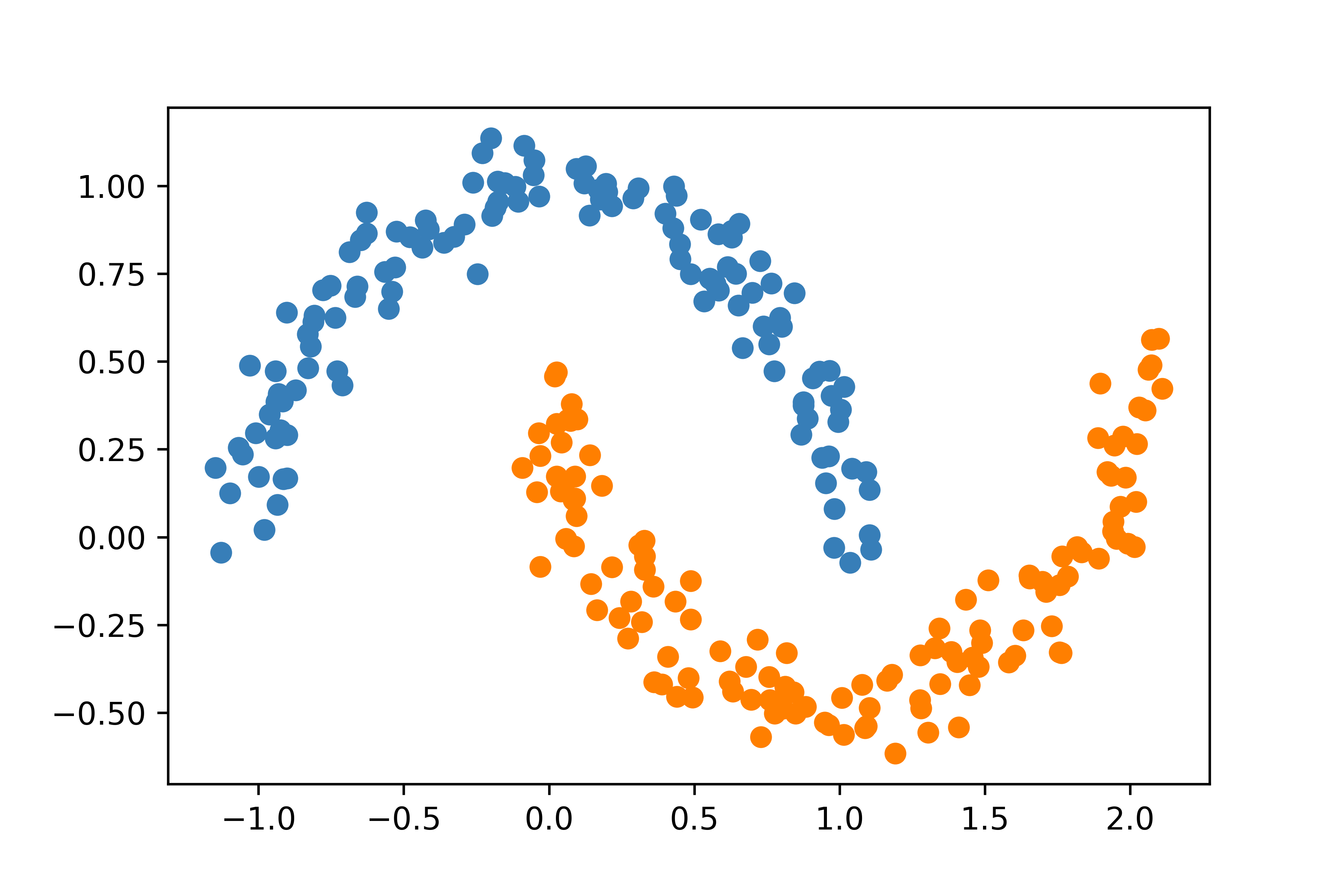}}
  \subfigure[]{
  \label{Fig.Result1.3}
  \includegraphics[width=0.32\textwidth]{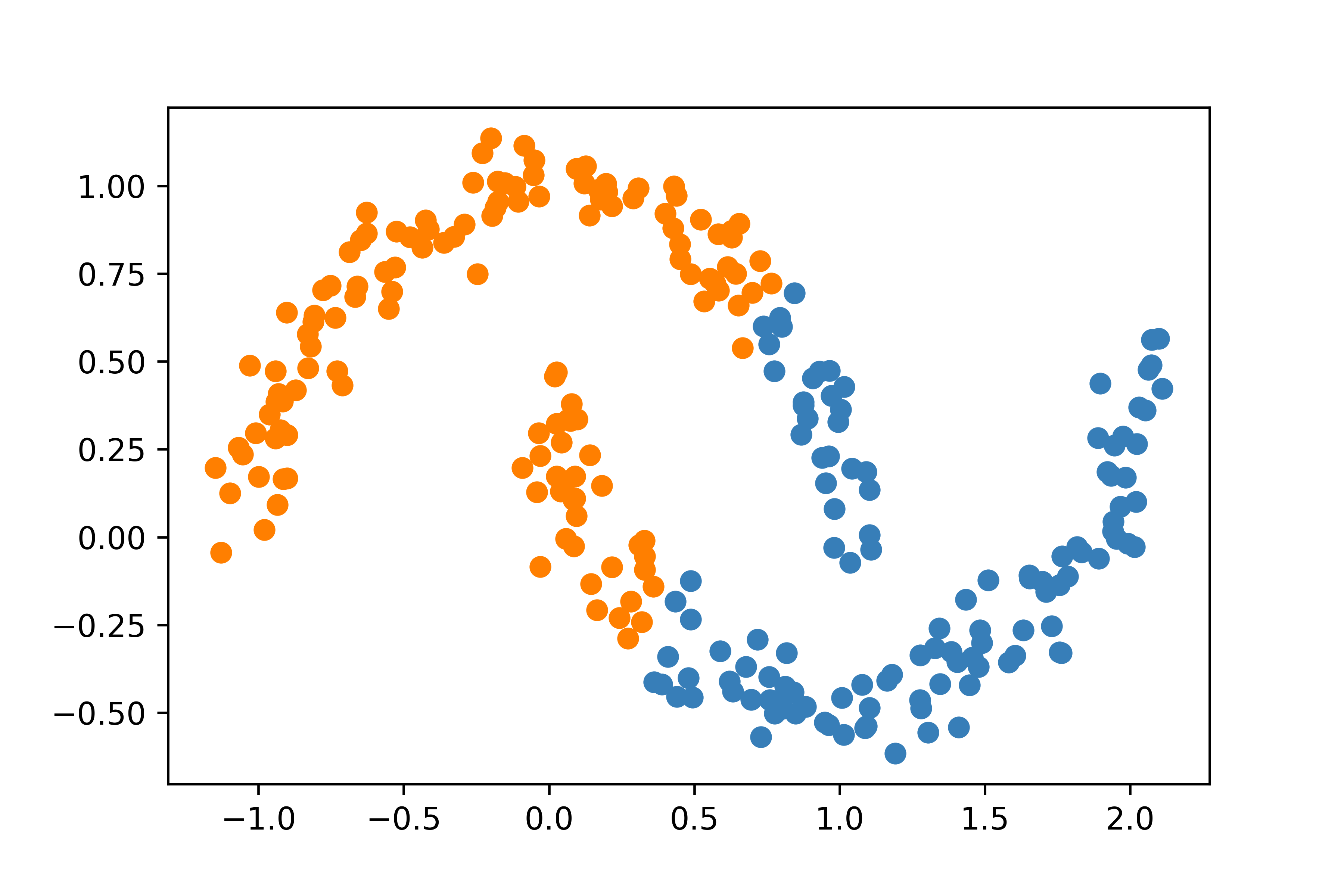}}
  \caption{(a) an illustrative image describing data set $D_1$ containing $256$ points and the corresponding $8$-nearest neighbor graph: two points are connected if they are an $8$-nearest neighbor to each other; (b) for $\tilde\lambda=\frac{1}{2^9}$, the clustering results for $D_1$ using our quantum spectral clustering algorithm and (c) the clustering result using classical $k$-means with $k=2$. The performance of our method for $D_1$ is much better than $k$-means.}
  \label{Result1}
\end{figure*}
\begin{figure*}[!t]
  \centering
  \subfigure[]{
  \label{Fig.Result2.1}
  \includegraphics[width=0.32\textwidth]{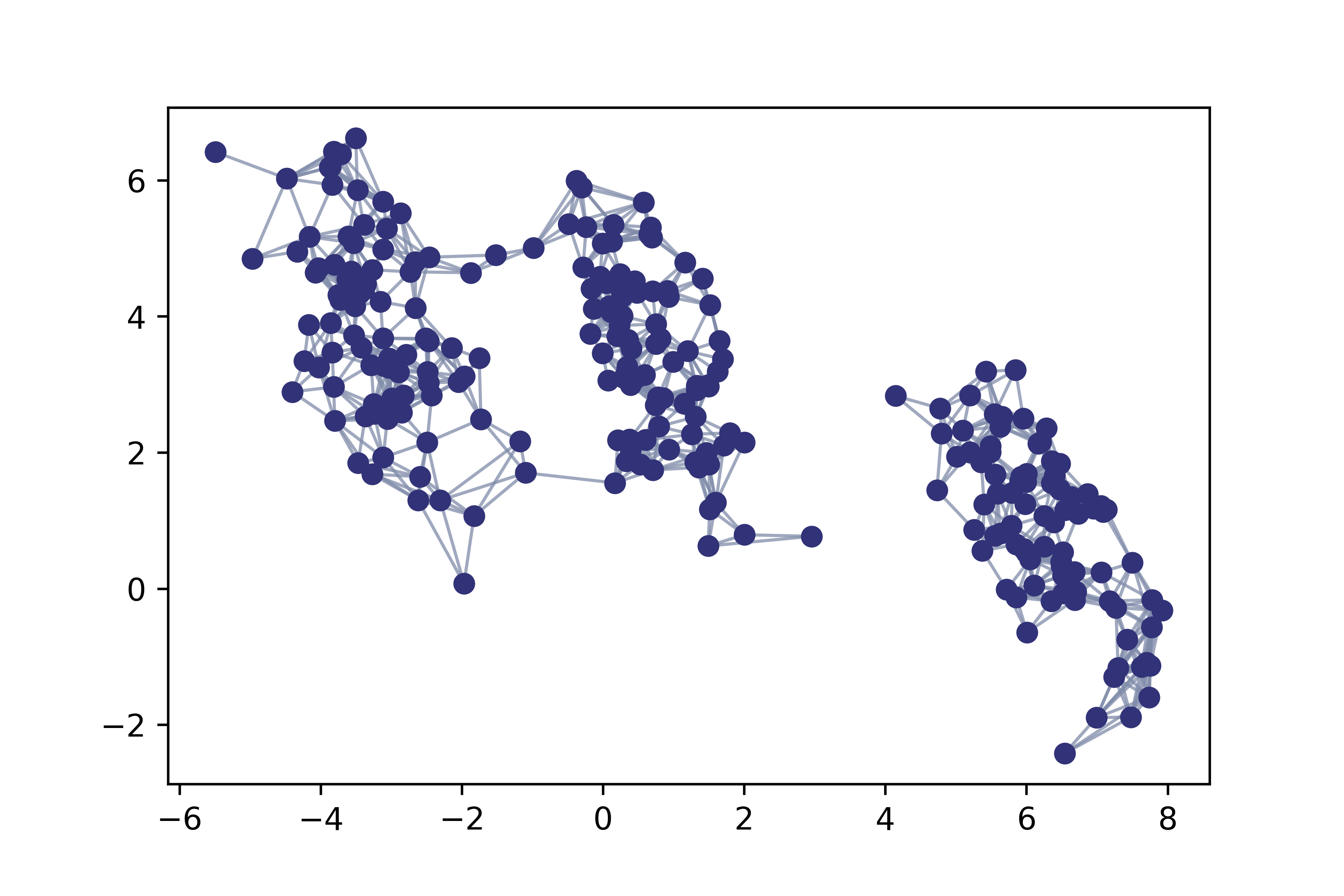}}
  \subfigure[]{
  \label{Fig.Result2.2}
  \includegraphics[width=0.32\textwidth]{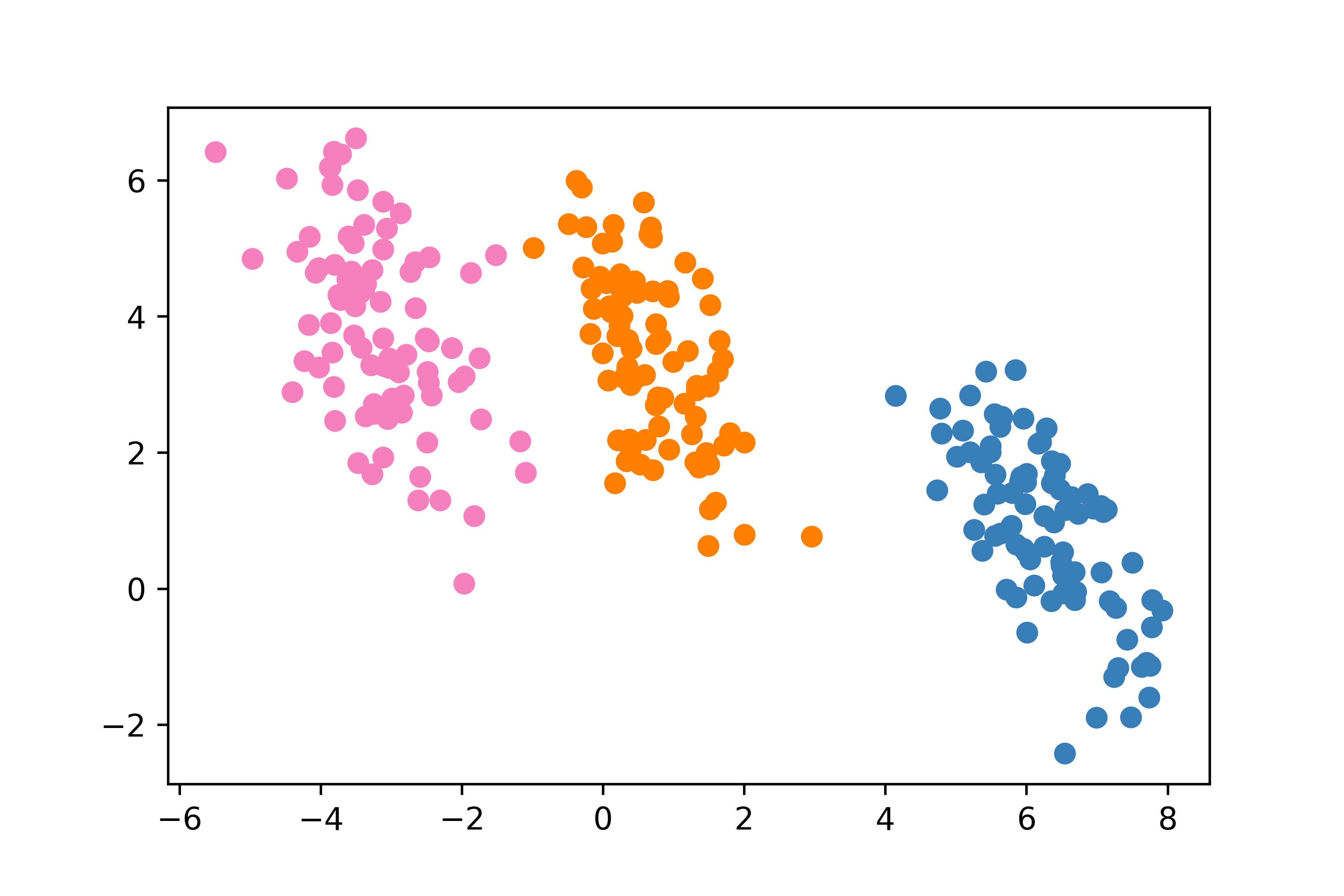}}
  \subfigure[]{
  \label{Fig.Result2.3}
  \includegraphics[width=0.32\textwidth]{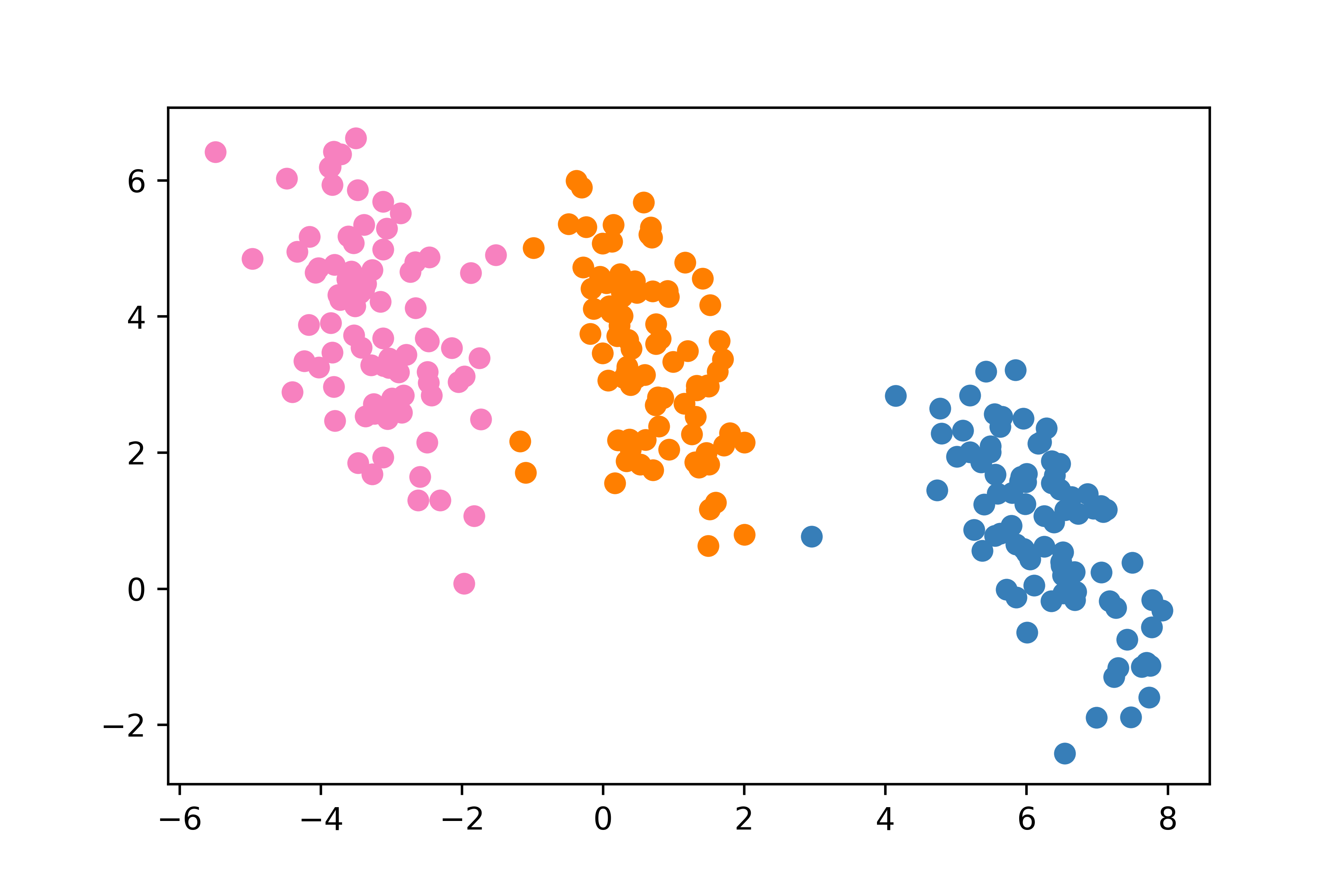}}
  \caption{(a) an illustrative image describing data set $D_2$ containing 256 points and the corresponding $8$-nearest neighbor graph: two points are connected if they are an $8$-nearest neighbor to each other; (b) for $\tilde\lambda=\frac{1}{2^9}$, the clustering results for $D_1$ using our quantum spectral clustering algorithm and (c) the clustering result using classical $k$-means with $k=3$. The performance of $k$-means is reasonably good, except for a few points incorrectly clustered.}
  \label{Result2}
\end{figure*}
To demonstrate how to implement our quantum spectral clustering proposal in solving specific spectral clustering problems, we apply it to two typical problems that are often used to benchmark the performance of clustering algorithms. The first example is a data set $D_1=\{\vec{v}^{(i)}\}_{i=0}^{255}$ defined on a $2$-dimensional compact set, satisfying $\vec{v}^{(i)}=[v^{(i)}_1,v^{(i)}_2]^T$ with $v^{(i)}_1\in[-1,2]$ and $v^{(i)}_2\in[-0.5,1]$, as illustrated in Figure~\ref{Result1} (a). According to Algorithm 2, we first generate the $8$-nearest neighbor graph of $D$, which will be used to generate the $9$-sparse $L_{1}\in\mathbb{R}^{256\times 256}$. Here we choose $d=8$ which is sufficient to give a good final clustering. Then we choose the threshold $\tilde \lambda=\frac{1}{2^9}$, prepare the initial state, and apply quantum counting to find the number of eigenvalues smaller than $\tilde \lambda$ to be $k=2$. Next, we implement the quantum spectral clustering circuit in Figure~\ref{Fig.quantumFlow} to obtain the final state $\rho$ of the eigenstate register. Then we choose an initial guess of the clustering indicator matrix $X=X^{(0)}$, and use hill climbing algorithm to optimize $\langle M\rangle=\Tr(M\rho)$ over $X$, with $M\equiv XX^T$. Through iterative optimization, we finally reach a sub-optimal solution $X=X^*$, which gives a clustering partition $\{P_0,P_1\}$, satisfying $P_j=\{\vec{v}_i|X^*_{ij}\neq 0\}$, for $i=0,\cdots,255$ and $j=0,1$. The clustering outcome is shown in Figure~\ref{Result1} (b), with the two clusters well separated, demonstrating a good clustering result. 

The second example is a dataset $D_2$ with $256$ data points $\vec{v}^{(i)}=[v^{(i)}_1,v^{(i)}_2]^T$ satisfying $v^{(i)}_1\in[-6,8]$ and $v^{(i)}_2\in[-2,6]$ (Figure~\ref{Result2} (a)). Analogously to the first example, we choose $d=8$ and construct the $9$-sparse Laplacian matrix $L$. Then we choose $\tilde \lambda=\frac{1}{2^9}$, and run the quantum counting algorithm to find $k=3$. After that we implement our spectral clustering circuit and hill-climbing algorithm, and through optimization we find a sub-optimal $X=X^*$, corresponding to a clustering partition $\{P_0,P_1,P_2\}$. The result is shown in Figure~\ref{Result2} (b). 

In comparison, we have also plotted the clustering results for $D_1$ and $D_2$ using $k$-means method, where we have chosen $k=2$ and $k=3$ respectively for the two examples, as shown in Figure~\ref{Result1} (c) and Figure~\ref{Result2} (c).  From these figures, one can see that although the $k$-means result for the second example is reasonably good, with only a few points incorrectly clustered, it is definitely not good for the first example. Hence, for problems like the first example, spectral clustering is necessary and advantageous to $k$-means algorithm. 

\textit{Conclusion} --
In this article, we explore the possibility of constructing a new quantum spectral clustering algorithm without using QRAM, but an appropriate combination of Grover's search, quantum phase estimation and Hamiltonian simulation. The essential point of the method is how to efficiently solve the eigenvalue problem on a quantum circuit and how to make use of the circuit outcome to complete the clustering task. One crucial trick of the method is the initial preparation of the registers in a bipartite maximally entangled state. We also utilize quantum counting to calculate the actual value of the number of clusters. A sub-optimal clustering outcome can be obtained by optimizing the measurement outcome using hill climbing algorithm. The overall quantum complexity of our method demonstrates a speedup compared to the classical counterpart. Our method provides a new instance to demonstrate the advantage of a quantum processor in solving machine learning problems, and we hope these techniques we have developed in the work can be applicable to other interesting quantum computational problems as well.

\textit{Acknowledgments} --
The authors gratefully acknowledge the grant from National Key R\&D Program of China, Grant No.2018YFA0306703.
\bibliography{reference}

\end{document}